\RequirePackage{fix-cm}
\documentclass[twocolumn]{svjour3}                     % onecolumn (standard format)
\smartqed  % flush right qed marks, e.g. at end of proof
\usepackage{hyperref}
\usepackage{marginnote}
\usepackage{makeidx}

\usepackage{graphicx}
\usepackage{float}
\usepackage{mathptmx}      % use Times fonts if available on your TeX system
\usepackage{amsmath}
\usepackage{subcaption}
\usepackage[export]{adjustbox}
\usepackage{booktabs}
\usepackage{multirow}
\usepackage{todonotes}
\usepackage[tophrase={~to~},binary-units=true,separate-uncertainty=true,multi-part-units=single,allow-number-unit-breaks=true,retain-explicit-plus]{siunitx}
\usepackage{pgfplots}
\usepgfplotslibrary{units}
\usepgfplotslibrary{colormaps}
\pgfplotsset{
	compat=newest,
	colormap={physicsmap}{
		rgb255=(47,2,119)
		rgb255=(46,2,120)
		rgb255=(45,2,122)
		rgb255=(44,2,123)
		rgb255=(42,2,124)
		rgb255=(41,2,125)
		rgb255=(40,2,126)
		rgb255=(39,2,127)
		rgb255=(37,2,128)
		rgb255=(36,1,129)
		rgb255=(35,1,131)
		rgb255=(33,1,132)
		rgb255=(32,1,133)
		rgb255=(30,1,134)
		rgb255=(29,1,135)
		rgb255=(27,1,136)
		rgb255=(26,1,137)
		rgb255=(24,1,138)
		rgb255=(23,1,139)
		rgb255=(21,1,141)
		rgb255=(19,1,142)
		rgb255=(18,1,143)
		rgb255=(16,1,144)
		rgb255=(14,1,145)
		rgb255=(12,1,146)
		rgb255=(10,1,147)
		rgb255=(9,0,148)
		rgb255=(7,0,149)
		rgb255=(5,0,151)
		rgb255=(3,0,152)
		rgb255=(1,0,153)
		rgb255=(0,1,154)
		rgb255=(0,3,155)
		rgb255=(0,5,156)
		rgb255=(0,7,157)
		rgb255=(0,9,158)
		rgb255=(0,11,160)
		rgb255=(0,13,161)
		rgb255=(0,15,162)
		rgb255=(0,18,163)
		rgb255=(1,20,164)
		rgb255=(2,23,165)
		rgb255=(3,26,166)
		rgb255=(3,28,167)
		rgb255=(4,31,168)
		rgb255=(5,33,169)
		rgb255=(6,36,169)
		rgb255=(7,39,170)
		rgb255=(8,41,171)
		rgb255=(8,44,172)
		rgb255=(9,47,173)
		rgb255=(10,49,174)
		rgb255=(11,52,175)
		rgb255=(12,55,176)
		rgb255=(13,58,177)
		rgb255=(14,60,178)
		rgb255=(15,63,179)
		rgb255=(16,66,180)
		rgb255=(16,69,181)
		rgb255=(17,71,182)
		rgb255=(18,74,183)
		rgb255=(19,77,184)
		rgb255=(20,80,185)
		rgb255=(21,82,186)
		rgb255=(22,85,187)
		rgb255=(23,88,188)
		rgb255=(24,91,189)
		rgb255=(25,94,190)
		rgb255=(26,96,191)
		rgb255=(26,99,191)
		rgb255=(26,102,191)
		rgb255=(26,105,191)
		rgb255=(26,108,192)
		rgb255=(26,111,192)
		rgb255=(26,114,192)
		rgb255=(26,116,192)
		rgb255=(26,119,192)
		rgb255=(26,122,193)
		rgb255=(26,125,193)
		rgb255=(26,128,193)
		rgb255=(26,131,193)
		rgb255=(26,134,193)
		rgb255=(26,137,194)
		rgb255=(26,140,194)
		rgb255=(26,143,194)
		rgb255=(26,146,194)
		rgb255=(26,149,194)
		rgb255=(26,152,195)
		rgb255=(26,155,195)
		rgb255=(25,158,195)
		rgb255=(25,161,195)
		rgb255=(25,164,195)
		rgb255=(25,167,196)
		rgb255=(25,170,196)
		rgb255=(25,173,196)
		rgb255=(25,176,196)
		rgb255=(25,179,196)
		rgb255=(25,182,197)
		rgb255=(25,185,197)
		rgb255=(25,188,197)
		rgb255=(25,191,197)
		rgb255=(25,194,197)
		rgb255=(25,198,198)
		rgb255=(25,198,195)
		rgb255=(25,198,192)
		rgb255=(25,198,190)
		rgb255=(25,198,187)
		rgb255=(25,199,184)
		rgb255=(25,199,181)
		rgb255=(25,199,178)
		rgb255=(25,199,176)
		rgb255=(25,199,173)
		rgb255=(25,200,170)
		rgb255=(24,200,167)
		rgb255=(24,200,165)
		rgb255=(24,200,162)
		rgb255=(24,200,159)
		rgb255=(24,201,156)
		rgb255=(24,201,153)
		rgb255=(24,201,150)
		rgb255=(24,201,148)
		rgb255=(24,202,143)
		rgb255=(24,202,138)
		rgb255=(24,202,133)
		rgb255=(24,203,129)
		rgb255=(24,203,124)
		rgb255=(24,203,119)
		rgb255=(24,204,114)
		rgb255=(24,204,109)
		rgb255=(24,204,104)
		rgb255=(23,205,99)
		rgb255=(23,205,94)
		rgb255=(23,205,89)
		rgb255=(23,206,84)
		rgb255=(23,206,79)
		rgb255=(23,206,74)
		rgb255=(23,207,69)
		rgb255=(23,207,64)
		rgb255=(23,207,59)
		rgb255=(23,208,54)
		rgb255=(23,208,49)
		rgb255=(23,208,44)
		rgb255=(23,209,39)
		rgb255=(22,209,33)
		rgb255=(22,209,28)
		rgb255=(22,210,23)
		rgb255=(27,210,22)
		rgb255=(32,210,22)
		rgb255=(37,211,22)
		rgb255=(42,211,22)
		rgb255=(48,211,22)
		rgb255=(53,212,22)
		rgb255=(58,212,22)
		rgb255=(63,212,22)
		rgb255=(69,213,22)
		rgb255=(74,213,21)
		rgb255=(79,213,21)
		rgb255=(85,214,21)
		rgb255=(90,214,21)
		rgb255=(96,214,21)
		rgb255=(101,215,21)
		rgb255=(107,215,21)
		rgb255=(112,215,21)
		rgb255=(118,216,21)
		rgb255=(123,216,21)
		rgb255=(129,216,21)
		rgb255=(134,217,21)
		rgb255=(140,217,20)
		rgb255=(145,217,20)
		rgb255=(148,217,20)
		rgb255=(151,217,20)
		rgb255=(153,218,20)
		rgb255=(156,218,20)
		rgb255=(159,218,20)
		rgb255=(161,218,20)
		rgb255=(164,218,20)
		rgb255=(166,218,20)
		rgb255=(169,219,20)
		rgb255=(172,219,20)
		rgb255=(174,219,20)
		rgb255=(177,219,20)
		rgb255=(180,219,20)
		rgb255=(182,219,20)
		rgb255=(185,219,20)
		rgb255=(188,220,20)
		rgb255=(190,220,20)
		rgb255=(193,220,20)
		rgb255=(196,220,20)
		rgb255=(198,220,19)
		rgb255=(201,220,19)
		rgb255=(204,221,19)
		rgb255=(206,221,19)
		rgb255=(210,221,19)
		rgb255=(213,221,19)
		rgb255=(216,221,19)
		rgb255=(220,221,19)
		rgb255=(222,220,19)
		rgb255=(222,217,19)
		rgb255=(222,214,19)
		rgb255=(222,212,19)
		rgb255=(222,209,19)
		rgb255=(223,206,19)
		rgb255=(223,203,19)
		rgb255=(223,200,19)
		rgb255=(223,197,19)
		rgb255=(223,194,19)
		rgb255=(223,191,19)
		rgb255=(224,188,19)
		rgb255=(224,185,18)
		rgb255=(224,182,18)
		rgb255=(224,178,18)
		rgb255=(224,175,18)
		rgb255=(225,172,18)
		rgb255=(225,169,18)
		rgb255=(225,166,18)
		rgb255=(225,163,18)
		rgb255=(225,160,18)
		rgb255=(225,157,18)
		rgb255=(226,154,18)
		rgb255=(226,151,18)
		rgb255=(226,148,18)
		rgb255=(226,145,18)
		rgb255=(226,141,18)
		rgb255=(227,138,18)
		rgb255=(227,135,18)
		rgb255=(227,132,18)
		rgb255=(227,129,17)
		rgb255=(227,126,17)
		rgb255=(228,122,17)
		rgb255=(228,119,17)
		rgb255=(228,116,17)
		rgb255=(228,113,17)
		rgb255=(228,110,17)
		rgb255=(228,106,17)
		rgb255=(229,103,17)
		rgb255=(229,100,17)
		rgb255=(229,97,17)
		rgb255=(229,94,17)
		rgb255=(229,90,17)
		rgb255=(230,87,17)
		rgb255=(230,84,17)
		rgb255=(230,80,17)
		rgb255=(230,77,17)
		rgb255=(230,74,17)
		rgb255=(230,71,16)
		rgb255=(231,67,16)
		rgb255=(231,64,16)
		rgb255=(231,61,16)
		rgb255=(231,57,16)
		rgb255=(231,54,16)
		rgb255=(232,51,16)
		rgb255=(232,47,16)
		rgb255=(232,44,16)
		rgb255=(232,41,16)
		rgb255=(232,37,16)
		rgb255=(233,34,16)
	},
	colormap={dosemap}{
		rgb255=(255,255,255)
		rgb255=(251,252,253)
		rgb255=(247,249,252)
		rgb255=(244,246,251)
		rgb255=(240,244,250)
		rgb255=(236,241,249)
		rgb255=(233,238,248)
		rgb255=(229,236,247)
		rgb255=(225,233,246)
		rgb255=(222,230,245)
		rgb255=(218,227,244)
		rgb255=(214,225,243)
		rgb255=(211,222,242)
		rgb255=(207,219,240)
		rgb255=(203,217,239)
		rgb255=(200,214,238)
		rgb255=(196,211,237)
		rgb255=(192,209,236)
		rgb255=(189,206,235)
		rgb255=(185,203,234)
		rgb255=(181,200,233)
		rgb255=(178,198,232)
		rgb255=(174,195,231)
		rgb255=(170,192,230)
		rgb255=(167,190,229)
		rgb255=(163,187,228)
		rgb255=(159,184,226)
		rgb255=(156,182,225)
		rgb255=(152,179,224)
		rgb255=(148,176,223)
		rgb255=(145,173,222)
		rgb255=(141,171,221)
		rgb255=(138,168,220)
		rgb255=(134,165,219)
		rgb255=(130,163,218)
		rgb255=(127,160,217)
		rgb255=(123,157,216)
		rgb255=(119,154,215)
		rgb255=(116,152,214)
		rgb255=(112,149,212)
		rgb255=(108,146,211)
		rgb255=(105,144,210)
		rgb255=(101,141,209)
		rgb255=(97,138,208)
		rgb255=(94,136,207)
		rgb255=(90,133,206)
		rgb255=(86,130,205)
		rgb255=(83,127,204)
		rgb255=(79,125,203)
		rgb255=(75,122,202)
		rgb255=(72,119,201)
		rgb255=(68,117,200)
		rgb255=(64,114,198)
		rgb255=(61,111,197)
		rgb255=(57,109,196)
		rgb255=(53,106,195)
		rgb255=(50,103,194)
		rgb255=(46,100,193)
		rgb255=(42,98,192)
		rgb255=(39,95,191)
		rgb255=(35,92,190)
		rgb255=(31,90,189)
		rgb255=(28,87,188)
		rgb255=(24,84,187)
		rgb255=(22,85,187)
		rgb255=(23,88,188)
		rgb255=(24,91,189)
		rgb255=(25,94,190)
		rgb255=(26,96,191)
		rgb255=(26,99,191)
		rgb255=(26,102,191)
		rgb255=(26,105,191)
		rgb255=(26,108,192)
		rgb255=(26,111,192)
		rgb255=(26,114,192)
		rgb255=(26,116,192)
		rgb255=(26,119,192)
		rgb255=(26,122,193)
		rgb255=(26,125,193)
		rgb255=(26,128,193)
		rgb255=(26,131,193)
		rgb255=(26,134,193)
		rgb255=(26,137,194)
		rgb255=(26,140,194)
		rgb255=(26,143,194)
		rgb255=(26,146,194)
		rgb255=(26,149,194)
		rgb255=(26,152,195)
		rgb255=(26,155,195)
		rgb255=(25,158,195)
		rgb255=(25,161,195)
		rgb255=(25,164,195)
		rgb255=(25,167,196)
		rgb255=(25,170,196)
		rgb255=(25,173,196)
		rgb255=(25,176,196)
		rgb255=(25,179,196)
		rgb255=(25,182,197)
		rgb255=(25,185,197)
		rgb255=(25,188,197)
		rgb255=(25,191,197)
		rgb255=(25,194,197)
		rgb255=(25,198,198)
		rgb255=(25,198,195)
		rgb255=(25,198,192)
		rgb255=(25,198,190)
		rgb255=(25,198,187)
		rgb255=(25,199,184)
		rgb255=(25,199,181)
		rgb255=(25,199,178)
		rgb255=(25,199,176)
		rgb255=(25,199,173)
		rgb255=(25,200,170)
		rgb255=(24,200,167)
		rgb255=(24,200,165)
		rgb255=(24,200,162)
		rgb255=(24,200,159)
		rgb255=(24,201,156)
		rgb255=(24,201,153)
		rgb255=(24,201,150)
		rgb255=(24,201,148)
		rgb255=(24,202,143)
		rgb255=(24,202,138)
		rgb255=(24,202,133)
		rgb255=(24,203,129)
		rgb255=(24,203,124)
		rgb255=(24,203,119)
		rgb255=(24,204,114)
		rgb255=(24,204,109)
		rgb255=(24,204,104)
		rgb255=(23,205,99)
		rgb255=(23,205,94)
		rgb255=(23,205,89)
		rgb255=(23,206,84)
		rgb255=(23,206,79)
		rgb255=(23,206,74)
		rgb255=(23,207,69)
		rgb255=(23,207,64)
		rgb255=(23,207,59)
		rgb255=(23,208,54)
		rgb255=(23,208,49)
		rgb255=(23,208,44)
		rgb255=(23,209,39)
		rgb255=(22,209,33)
		rgb255=(22,209,28)
		rgb255=(22,210,23)
		rgb255=(27,210,22)
		rgb255=(32,210,22)
		rgb255=(37,211,22)
		rgb255=(42,211,22)
		rgb255=(48,211,22)
		rgb255=(53,212,22)
		rgb255=(58,212,22)
		rgb255=(63,212,22)
		rgb255=(69,213,22)
		rgb255=(74,213,21)
		rgb255=(79,213,21)
		rgb255=(85,214,21)
		rgb255=(90,214,21)
		rgb255=(96,214,21)
		rgb255=(101,215,21)
		rgb255=(107,215,21)
		rgb255=(112,215,21)
		rgb255=(118,216,21)
		rgb255=(123,216,21)
		rgb255=(129,216,21)
		rgb255=(134,217,21)
		rgb255=(140,217,20)
		rgb255=(145,217,20)
		rgb255=(148,217,20)
		rgb255=(151,217,20)
		rgb255=(153,218,20)
		rgb255=(156,218,20)
		rgb255=(159,218,20)
		rgb255=(161,218,20)
		rgb255=(164,218,20)
		rgb255=(166,218,20)
		rgb255=(169,219,20)
		rgb255=(172,219,20)
		rgb255=(174,219,20)
		rgb255=(177,219,20)
		rgb255=(180,219,20)
		rgb255=(182,219,20)
		rgb255=(185,219,20)
		rgb255=(188,220,20)
		rgb255=(190,220,20)
		rgb255=(193,220,20)
		rgb255=(196,220,20)
		rgb255=(198,220,19)
		rgb255=(201,220,19)
		rgb255=(204,221,19)
		rgb255=(206,221,19)
		rgb255=(210,221,19)
		rgb255=(213,221,19)
		rgb255=(216,221,19)
		rgb255=(220,221,19)
		rgb255=(222,220,19)
		rgb255=(222,217,19)
		rgb255=(222,214,19)
		rgb255=(222,212,19)
		rgb255=(222,209,19)
		rgb255=(223,206,19)
		rgb255=(223,203,19)
		rgb255=(223,200,19)
		rgb255=(223,197,19)
		rgb255=(223,194,19)
		rgb255=(223,191,19)
		rgb255=(224,188,19)
		rgb255=(224,185,18)
		rgb255=(224,182,18)
		rgb255=(224,178,18)
		rgb255=(224,175,18)
		rgb255=(225,172,18)
		rgb255=(225,169,18)
		rgb255=(225,166,18)
		rgb255=(225,163,18)
		rgb255=(225,160,18)
		rgb255=(225,157,18)
		rgb255=(226,154,18)
		rgb255=(226,151,18)
		rgb255=(226,148,18)
		rgb255=(226,145,18)
		rgb255=(226,141,18)
		rgb255=(227,138,18)
		rgb255=(227,135,18)
		rgb255=(227,132,18)
		rgb255=(227,129,17)
		rgb255=(227,126,17)
		rgb255=(228,122,17)
		rgb255=(228,119,17)
		rgb255=(228,116,17)
		rgb255=(228,113,17)
		rgb255=(228,110,17)
		rgb255=(228,106,17)
		rgb255=(229,103,17)
		rgb255=(229,100,17)
		rgb255=(229,97,17)
		rgb255=(229,94,17)
		rgb255=(229,90,17)
		rgb255=(230,87,17)
		rgb255=(230,84,17)
		rgb255=(230,80,17)
		rgb255=(230,77,17)
		rgb255=(230,74,17)
		rgb255=(230,71,16)
		rgb255=(231,67,16)
		rgb255=(231,64,16)
		rgb255=(231,61,16)
		rgb255=(231,57,16)
		rgb255=(231,54,16)
		rgb255=(232,51,16)
		rgb255=(232,47,16)
		rgb255=(232,44,16)
		rgb255=(232,41,16)
		rgb255=(232,37,16)
		rgb255=(233,34,16)
	}
}
\newcommand{\longaxis}{longitudinal}
\newcommand{\absdiv}[1]{%
  \par\addvspace{.5\baselineskip}% adjust to suit
  \noindent\textbf{#1}\quad\ignorespaces
}
\definecolor{red}{rgb}{0.8,0,0}
\definecolor{blue}{rgb}{0,0,0.8}
\definecolor{green}{rgb}{0,0.4,0}
\newcommand{\change}[2]{}
\newcommand{\lchange}[2]{}
\newcommand{\changedtext}{}
\newcommand{\changed}[3]{#3}

\begin{document}

\title{Pitfalls in interventional X-ray organ dose assessment---combined experimental and computational phantom study: application to prostatic artery embolization}

\titlerunning{Pitfalls in interventional X-ray organ dose assessment}        % if too long for running head

\author{
    Philipp~Roser$^{1,3}$, Annette~Birkhold$^2$, Xia~Zhong$^2$, Philipp~Ochs$^2$, Elizaveta~Stepina$^2$, Markus~Kowarschik$^2$,  Rebecca~Fahrig$^2$, Andreas~Maier$^{1,3}$
}

\authorrunning{Roser et al.} % if too long for running head

\institute{
    $^1$Pattern Recognition Lab, FAU Erlangen-N\"urnberg, Erlangen, Germany\\
    $^2$Siemens Healthcare GmbH, Forchheim, Germany\\
    $^3$Erlangen Graduate School in Advanced Optical Technologies (SAOT), Erlangen, Germany\\
    \email{philipp.roser@fau.de}\\
}

\date{Received: 18 February 2019 / Accepted: 19 July 2019}

\maketitle

\pagenumbering{arabic}
\begin{abstract}
\hspace{0pt}\vspace{0.5\baselineskip}
\absdiv{Purpose}
With X-ray radiation protection and dose management constantly gaining interest in interventional radiology, novel procedures often undergo prospective dose studies using anthropomorphic phantoms to determine expected reference organ-equivalent dose values. 
Due to inherent uncertainties, such as impact of exact patient positioning, generalized geometry of the phantoms, limited dosimeter positioning options, and composition of tissue-equivalent materials, these dose values might not allow for patient-specific risk assessment.
Therefore, first the aim of this study is to quantify the influence of these parameters on local X-ray dose to evaluate their relevance in the assessment of patient-specific organ doses. Second, this knowledge further enables validating a simulation approach, which allows employing physiological material models and patient-specific geometries.
\absdiv{Methods}
Phantom dosimetry experiments using MOSFET dosimeters were conducted reproducing imaging scenarios in prostatic arterial embolization (PAE). Associated organ-equivalent dose of prostate, bladder, colon and skin was determined. Dose deviation induced by possible small displacements of the patient was reproduced by moving the X-ray source. Dose deviation induced by geometric and material differences was investigated by analyzing two different commonly used phantoms.
We reconstructed the experiments using Monte Carlo (MC) simulations, a reference male geometry, and different material properties to validate simulations and experiments against each other.
\absdiv{Results}
Overall, MC simulated organ dose values are in accordance with the measured ones for the majority of cases.
Marginal displacements of X-ray source relative to the phantoms lead to deviations of \SIrange{6}{135}{\percent} in organ dose values, while skin dose remains relatively constant.
Regarding the impact of phantom material composition, underestimation of internal organ dose values by \SIrange{12}{20}{\percent} is prevalent in all simulated phantoms.
Skin dose, however, can be estimated with low deviation of \SIrange{1}{8}{\percent} at least for two materials.
\absdiv{Conclusions}
Prospective reference dose studies might not extend to precise patient-specific dose assessment. 
Therefore online organ dose assessment tools, based on advanced patient modeling and MC methods are desirable. 
\end{abstract}

\section{Introduction}
\label{sec:introduction}

Over the last decades, the number of fluoroscopically-guided interventions (FGI) increased considerably, putting a stronger focus on radiation protection, management, and safety for both patients and radiologists.
In contrast to diagnostic radiology, complex FGIs, such as prostatic artery embolization (PAE), may require long fluoroscopic times or high-quality images, leading to increased radiation exposure \cite{Falco:18}, and therefore considerable deterministic and stochastic risks induced by the ionizing radiation.

% First, deterministic risks include skin injuries, also referred to as radiation dermatitis, and other observable adverse effects such as hair loss \cite{Balter:10,Wagner:99,Koenig:01,Dincan:02}.
First, \changed{M3.1}{\link{R3.1}}{in general} deterministic risks \changedtext{may} include skin injuries, also referred to as radiation dermatitis, and other observable adverse effects such as hair loss \changedtext{during neuro-interventional procedures} \cite{Balter:10,Wagner:99,Koenig:01,Dincan:02}.
Deterministic consequences are usually assessed based on the entrance peak skin dose (PSD), the highest irradiation accumulated on the patient's skin.
The PSD can be (1) measured directly using radiochromic film or comparable dosimeters, or (2) estimated indirectly based on the air kerma at the interventional reference point (K\textsubscript{IRP}) and dose-area-products (DAP), which are typically measured in \si{\gray} and \si{\gray\square\cm}, respectively \cite{Balter:06}.

Second, stochastic risks mainly describe the increased probability to develop some form of cancer.
This risk is quantified using either effective dose ($E$) or tissue-specific equivalent doses ($H_T$).
Both are usually measured in \si{\sievert}, however, since X-rays only include photon and electron interactions, $H_T$ values can also be given with respect to \si{\gray}.
While $E$ is typically correlated linearly with the measured DAP, its usage as only risk estimator is dissuaded since it is determined with respect to a reference human model and a specific, reproducible procedure, which is not given in interventional radiology necessarily \cite{Falco:18,Mattsson:16}.
In general, the direct measurement of $E$ and $H_T$ is not feasible in-vivo and only possible in phantom or simulation studies.

% While most FGIs and associated dose deposition are justified by increased mortality due to not treating the present condition, other interventions are mainly performed to increase the life quality of the patient, such as PAE.
\changed{M3.2}{\link{R3.2}}{
Most FGIs are performed to treat potentially fatal conditions, and resulting high dose values are therefore justified evidently.
However, radiation doses applied during procedures aimed at improving quality of life mainly, such as during PAE, have to be weighed against the benefit of the procedure more strictly.
}
Recently, PAE gained a notable increase in popularity as a primary treatment for benign prostatic hyperplasia \cite{Carnevale:10,Bagla:13,Christidis:18}.
Since PAE heavily relies on fluoroscopic guidance, digital subtraction angiography (DSA), and cone-beam computed tomography (CBCT), patients' skin and effective dose, as well as the radiation exposure of the treating medical staff are serious concerns and may vary considerably between different procedures \cite{Garzon:16,Laborda:15}.

So far, the patient exposure during PAEs has mostly been assessed based on the PSD, measured directly using radio\-chromic film \cite{Garzon:16,Andrade:17}, or based on K\textsubscript{IRP} and DAP \cite{Tanaka:17,Gao:14,Pisco:16}.
At the time of this study, there has been only one (phantom) study focused on patient $E$ or $H_T$ values published \cite{Andrade:17}.
However, it is challenging to generalize (phantom) study results to specific patients and procedures as there are multiple sources of uncertainty involved.
For instance, patient anatomy and placement may vary notably between procedures, leading to increased or reduced organ dose for specific irradiated organs.
In addition, FGIs are less reproducible as diagnostic CTs, as they might require patient-specific workflow adaptations over the course of the intervention.
This is further underlined in the example of PAE, where large deviations in fluoroscopic time, number of DSA images, usage of CBCT, and overall PSD have been reported \cite{Andrade:17}.

It still remains questionable, whether predetermined expected reference $E$ or $H_T$ values are sufficient to pro-actively assess and plan radiation exposure online.
Instead, employing online or even retrospective Monte Carlo (MC) simulation of the radiation transport inside the patient seem attractive to actually estimate patient- and procedure-specific dose values that allow for the individual risk assessment.
Although MC methods usually introduce high computation latency, there exist successful implementations making use of GPU acceleration \cite{Badal:09,Bert:13}.
To this end, we combine empiric dose measurements using dedicated anthropomorphic phantoms, and MC simulations of the same experimental setup to estimate organ doses of directly irradiated organs.
By varying input parameters, such as material composition or patient placement, the impact of small uncertainties can be quantified, underlining the need for individual and specific dose estimation.

\section{Materials and Methods}
\label{sec:methods}

\begin{figure*}
	\centering
	\subcaptionbox{Section 34.\label{fig:cirs-atom-34}}
    {\includegraphics[width=0.2\linewidth]{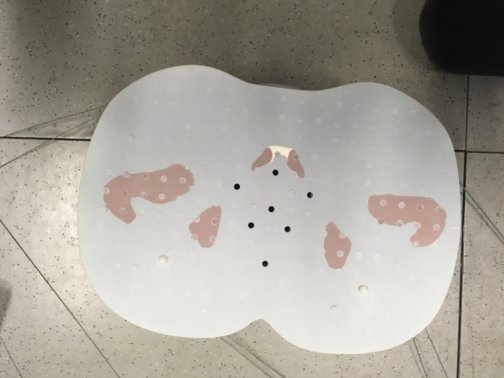}}
	\subcaptionbox{Section 35.\label{fig:cirs-atom-35}}
    {\includegraphics[width=0.2\linewidth]{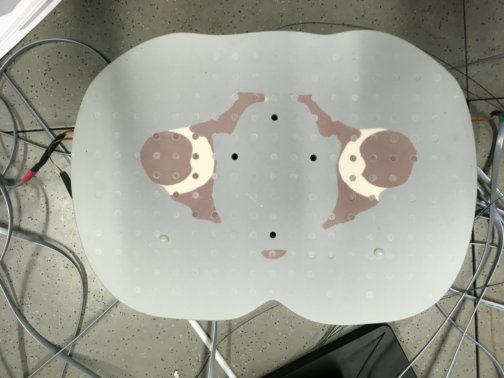}}
	\subcaptionbox{Skin entry.\label{fig:cirs-atom-entry}}
    {\includegraphics[width=0.2\linewidth]{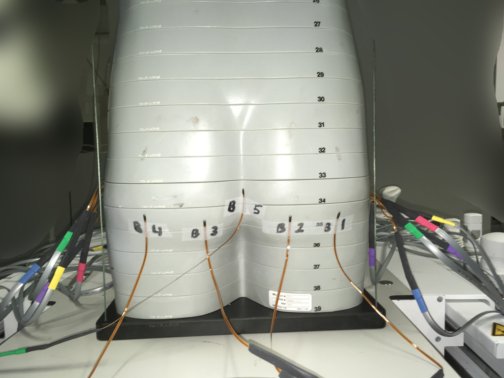}}
	\subcaptionbox{Skin exit.\label{fig:cirs-atom-exit}}
    {\includegraphics[width=0.2\linewidth]{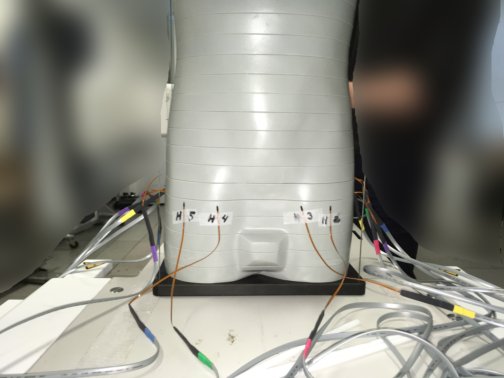}}   
    \subcaptionbox{X-ray.\label{fig:ca-roi}}{\includegraphics[width=0.116\textwidth]{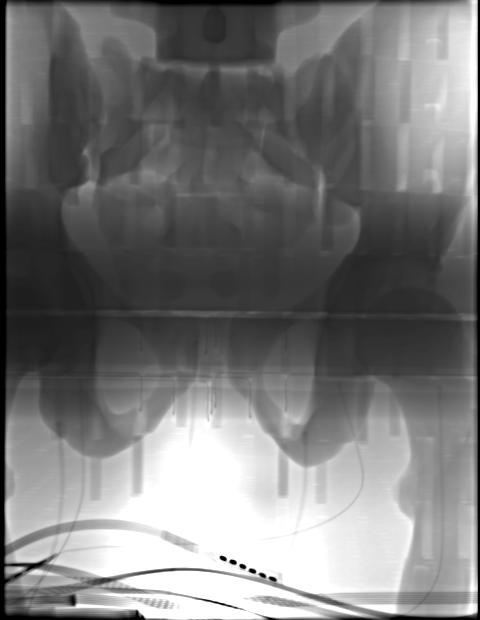}} 
    \caption{MOSFET dosimeter equipping and exemplary time-averaged acquisition of the CA phantom.}
    \label{fig:cirs-atom}
	\subcaptionbox{Section 33.\label{fig:alderson-rando-33}}
    {\includegraphics[width=0.2\linewidth]{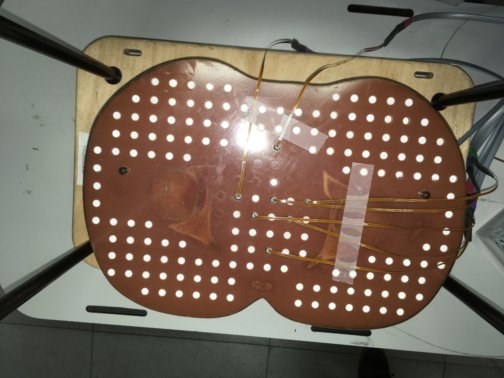}}
	\subcaptionbox{Section 34.\label{fig:alderson-rando-34}}
    {\includegraphics[width=0.2\linewidth,angle=180,origin=c]{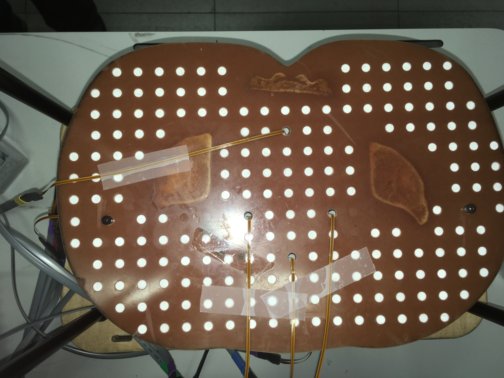}}
	\subcaptionbox{Skin entry.\label{fig:alderson-rando-entry}}
    {\includegraphics[width=0.2\linewidth]{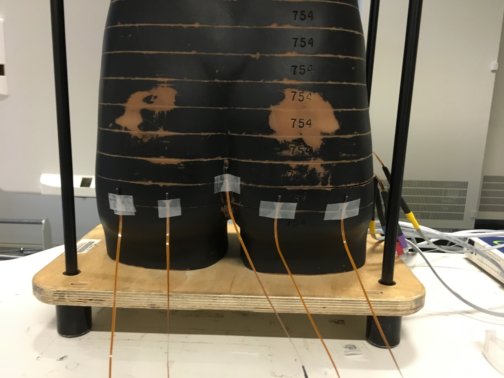}}
	\subcaptionbox{Skin exit.\label{fig:alderson-rando-exit}}
    {\includegraphics[width=0.2\linewidth]{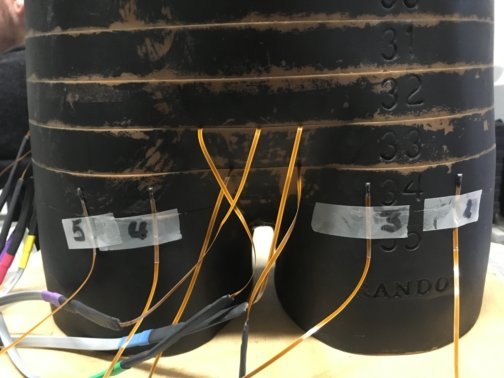}} 
    \subcaptionbox{X-ray.\label{fig:ar-roi}}{\includegraphics[width=0.116\textwidth]{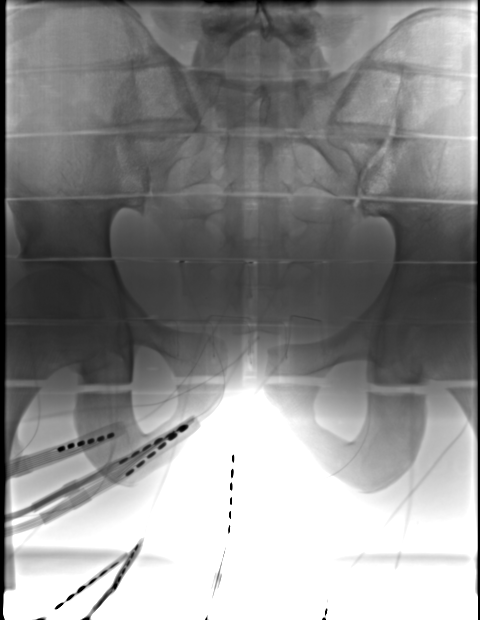}}       
    \caption{MOSFET dosimeter equipping and exemplary time-averaged acquisition of the AR phantom.}
    \label{fig:alderson-rando}
\end{figure*}

\begin{figure}
	\centering
    
	\subcaptionbox{Placement of the CA phantom.\label{fig:setup-ca}}
    {\includegraphics[width=0.45\linewidth]{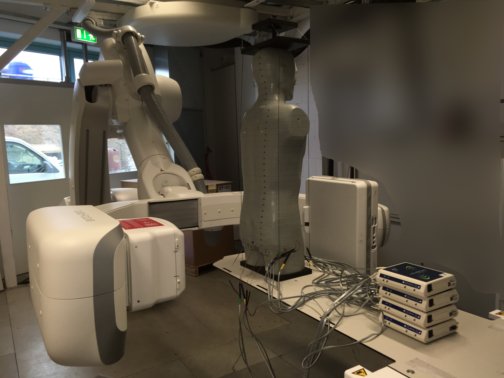}}
    \subcaptionbox{Placement of the AR phantom.\label{fig:setup-ar}}
    {\includegraphics[width=0.45\linewidth]{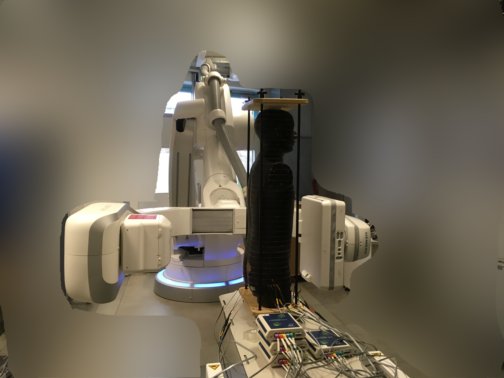}}
    
    \subcaptionbox{Central interface.\label{fig:setup-interface}}
    {\includegraphics[width=0.3\linewidth]{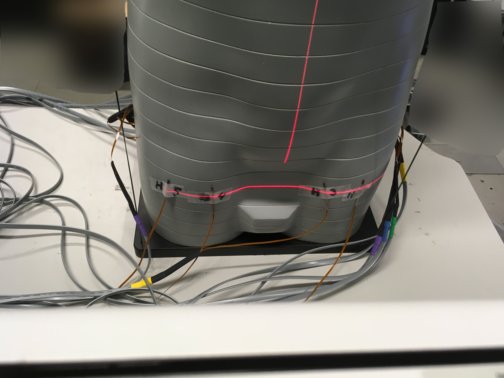}}
    \subcaptionbox{Interface above.\label{fig:setup-interface+}}
    {\includegraphics[width=0.3\linewidth]{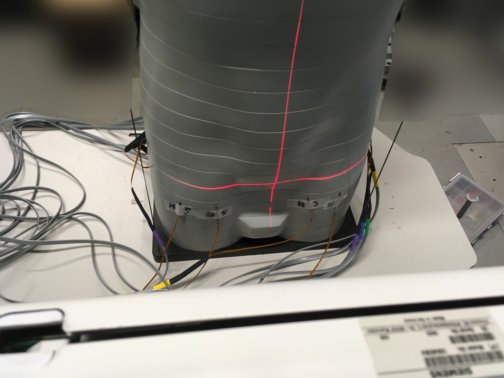}}
    \subcaptionbox{Interface below.\label{fig:setup-interface-}}
    {\includegraphics[width=0.3\linewidth]{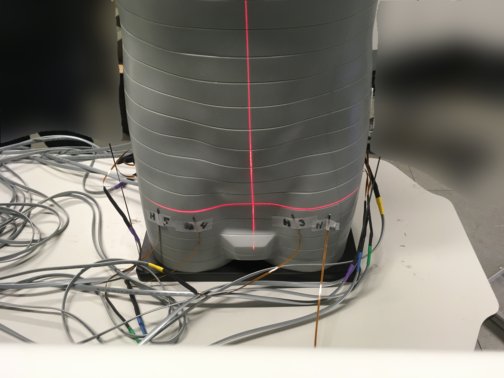}}
    
    \caption{Experiment setup and phantom placement.}
	\label{fig:setup}
\end{figure}

\subsection{Anthropomorphic phantoms}
To measure organ dose, we employed two anthropomorphic phantoms specifically designed for dosimetry purposes, the adult male RANDO phantom (AR), the predecessor of the Alderson Radiation Therapy phantom (RANDO, Radiology Support Devices, Inc., Long Beach, CA, USA), and the adult male ATOM phantom (CA) (ATOM Adult Male Model 701, Computerized Imaging Reference Systems, Inc., Norfolk, VA, USA). 

\subsubsection{Alderson RANDO (AR)}
The AR comprises 34, \SI{2.5}{\cm} thick, axial slices along the trunk and head, and a \SI{15}{\cm} thick slice of the legs, measuring \SI{97.5}{\cm} in total.
It consists of a human skeleton embedded in a synthetic isocyanate rubber compound, which is tissue-equivalent with respect to linear attenuation over the energy range typically used in external radiation therapy.
The AR phantom was shown to be water-equivalent with \SI{94}{\pm 1}{\percent} accuracy for \SI{70}{\kilo\volt}p spectra; however, this deviation has to be considered in diagnostic and interventional radiology dosimetry \cite{Shrimpton:81}.
Cavities to place dosimeters are uniformly distributed over each slice except where bone is located.
Unfortunately, the literature gives ambiguous information on the exact composition of the used rubber compound, stating (1) it is water-equivalent, (2) it is in accordance with the standard as proposed in the 44\textsuperscript{th} report of the International Commission on Radiation Units and Measurements (ICRU), or (3) the composition of elements by fraction of mass \cite{Shrimpton:81,White:78}.

\subsubsection{CIRS ATOM (CA)}
The CA comprises 39, \SI{2.5}{\cm} thick, axial slices along head, trunk and legs, measuring \SI{97.5}{\cm} in total.
It consists of averaged materials for soft/adipose, bone, lung (inhale) and brain tissue, which are tissue-equivalent with respect to linear attenuation in the range of \SI{50}{\kilo\eV} to \SI{15}{\mega\eV}.
In contrast to the AR phantom, the cavities to place dosimeters are uniformly distributed over each slice of the phantom and all present tissue types.
In addition, the manufacturer offers commercially available organ maps for each slice, which indicate the contours of internal radio-sensitive organs and potential dosimeter cavities to cover these.
The composition of each processed tissue-equivalent material is supplemented with respect to the contributing fraction of mass of each element. 

\subsection{Dosimeters}
For the assessment of organ dose, we used high-sensitivity metal oxide semiconductor field effect transistor (MOSFET) dosimeters TN 1002RD-H, equipped to a mobile reader (mobileMOSFET system, model TN-RD-70-W, Best Medical Canada Ltd., Ottawa, ON, Canada). 
The mobileMOSFET system consists of remote monitoring dose verification software, a Bluetooth wireless transceiver, and reader module that act as channel between MOSFET probes and software. 
Up to five MOSFETs can be connected to one reader.

Prior to the measurements, all MOSFET detectors were calibrated. 
For calibration purposes, each of used MOSFETs were irradiated with a specified dose level. 
The dose level was measured with an \SI{530}{\cubic\cm} ionization chamber (PM500-CII 52.8210, Capintec Inc., Ramsey, NJ, USA) connected to the Unidos dosimeter (PTW, Freiburg, Germany). 
The ionization chamber was calibrated by PTW accredited by the German National Accreditation Body (D-K 15059-01-00) as calibration laboratory in the \emph{Deutschen Kalibrierdienst} (German calibration service). 
The mobileMOSFET software calculates automatically the calibration factor as the ratio of the measured voltage value to the actual value of radiation dose for every single MOSFET probe.
\changed{M1.4}{\link{R1.4}}{According to the manufacturer, the uncertainty of the used MOSFET probes is below \SI{3}{\percent} at \SI{20}{\centi\gray}, for instance.
Extrapolating the manufacturer's data, the uncertainty is below \SI{11}{\percent} at \SI{20}{\milli\gray}.}

\subsection{Phantom equipping}
The MOSFET probes are distributed over two axial slices (section 34 and 35) of the CA phantom, and the entry and exit point of the X-ray on the phantom surface.
Following the manufacturer's organ maps, 11 measurement points (MP) cover prostate (3), bladder (6), and colon (2). 
The skin entrance dose is monitored by five dosimeters and the remainder of dosimeters monitors the exit dose.
The skin exit dose is measured mainly to validate MC simulation and experimental measurements against each other robustly.
Internally placed MOSFET dosimeters are placed in soft-tissue-equivalent holders to reduce the impact of air in the cavity, thus superseding corrective calculations based on a certain cavity theory.
All dosimeters are fixed using X-ray-transparent low-adhesive tape; the exact placement is shown in Figure~\ref{fig:cirs-atom}.

Accordingly, the AR phantom is equipped. 
Since overall anatomy and cavity spacing of both used phantoms differ slightly, we could only approximate the positioning of the MOSFET probes to match the CA.
Unfortunately, the section embedding the prostate is part of the bottom slice, which does not contain any cavities for dosimeters.
Therefore, the closest sections (section 33 and 34) and cavities are equipped with MOSFET dosimeters and the respective section identification numbers do not coincide with the sections of the CA phantom.
The exact probe placement is documented in Figure~\ref{fig:alderson-rando}.

\subsection{Experiment setup}
To simplify replicability and comparability between conducted, future, and possible replication phantom studies, the phantoms are placed upright and the X-ray system's C-arm is rotated by \SI{90}{\degree} as shown in Figure \ref{fig:setup-ca} and \ref{fig:setup-ar}, \changed{M3.3.1}{\link{R3.3}}{which is equivalent to anteroposterior position}.
\changedtext{
It is important to note that, during PAE, the X-ray system is usually rotated to an ipsilateral angle of \SIrange{25}{35}{\degree}.
However, to balance pertinence, reproducibility, and generalizability of this study, we decided to trivialize the experimental setup.
}
This approach also yields the advantage of eliminating uncertainties related to the multitude of available different interventional or surgery tables and mattresses.
During all conducted experiments, the same Artis zeego imaging system (Siemens Healthcare GmbH, Erlangen, Germany) is used.
The distance between X-ray tube and central \longaxis{} axis of both phantoms is \SI{800}{\mm}, and the overall source-to-image distance (SID) is \SI{1200}{\mm}.
The central X-ray is aligned with the interface of two axial sections using the built-in detector laser cross.
To investigate the impact of patient positioning and navigation of the C-arm on organ dose, three section interfaces are exposed for each phantom.
First, the interface between both sections that are equipped with MOSFET dosimeters is irradiated, followed by the interfaces \SI{25}{\mm} above and below of it as shown in Figure \ref{fig:setup-interface}, \ref{fig:setup-interface+}, and \ref{fig:setup-interface-}. These deviations are assumed to cover the possible patient position variations between different PAE procedures.

In order to secure sufficient exposure of all MOSFET probes, \SI{40}{\s} X-ray acquisitions with \num{30} frames per second are recorded.
The measurements are repeated three times for each image setting.
The peak tube voltage is \SI{70}{\kilo\volt} and neither additional filtration nor collimation is applied.
Automatic exposure control is deactivated resulting in a constant accumulated air kerma of \SI{40.8}{\milli\gray} per acquisition, which is directly monitored by an online ionization chamber.
After each angiography acquisition, the dosimeters are read out and the measured dose values for each MP are averaged.
With only photons and electrons being involved, the resulting organ-equivalent dose $H_T$ values are then given by the mean of all averaged MPs of the regarded organ, as the radiation weighting factor is $w_R = 1$ for these particles.
All measurements are taken on the same day using the same MOSFET devices and same equipping for both phantoms.
Figure \ref{fig:ca-roi} and \ref{fig:ar-roi} showcase time-averaged acquisitions for both investigated phantoms.

\subsection{Monte Carlo simulation}
% The simulation is implemented in a C++ application based on the general purpose MC toolkit Geant4 \cite{Agostinelli:03}, offering highly customizable and flexible interfaces to its kernel, allowing for arbitrary experiment configuration, particle tracking and scoring of quantities of dosimetric interest.
% Furthermore, Geant4 features the definition of custom material compounds and mixtures, enabling the accurate simulation of the AR and CA phantom materials. 
\changed{M1.1}{\link{R1.1}}{Geant4 \cite{Agostinelli:03} provides a customizable, flexible, and object-oriented interface to its open-source kernel written in C++ and was therefore used for the simulations.
        Although Geant4 stems from the high energy physics community, it is also well extensible to medical physics and features specific low-energy physical models \cite{Allison:16}. Four MC codes applicable to diagnostic X-ray dosimetry, namely EGSnrc, Geant4, MCNPX, and Penelope have been evaluated by the American Association of Physicists in Medicine \cite{Sechopoulos:15}.
	    To ensure validity of our particular Geant4-based application, we evaluated it using the test cases 1 and 2 as proposed by the American Association of Physicists in Medicine \cite{Sechopoulos:15}. No striking features were observed.}

\subsubsection{Digital phantom}
To employ the investigated phantoms in the MC simulation, we use the geometry of the reference voxel phantom Golem provided by the former Institute of Radiation Protection, which is now integrated into the  Institute of Innovative Radiotherapy\footnote{www.helmholtz-muenchen.de/en/institute-of-innovative-radiotherapy/index.html, accessed January 28\textsuperscript{th} 2019}.
The Golem phantom is an implementation of the male reference human as recommended by the International Commission on Radiological Protection (ICRP) \cite{ICRP:110}.
The \SI{176}{\cm} tall phantom consists of 220 slices with \num{256 x 256} voxels each, ranging from the vertex down to the toes.
The voxels have a volume of \SI{8 x 2.08 x 2.08}{\mm} and are segmented and labeled with respect to 122 organs and individual bones.

Four different voxel-wise material composition mappings are used to assign material properties to the associated labels. The first material composition represents the CA phantom.
Two material composition mappings reference the AR phantom in two different implementations, according to literature \cite{Shrimpton:81,White:78}.
First, one implementation follows the widely accepted material composition as proposed by White \cite{White:78}.
Second, the other implementation of the AR phantom (ARW) models soft and adipose tissue as water, which is a widely used simplification for soft and adipose tissue.
The fourth material mapping serves as reference mapping (R) and is modeled to resemble a living adult male, following the material specifications proposed by the ICRP standard.
It comprises adipose, soft, skin, brain, bone (cortical), muscle, and lung (inhale) tissue.
The skin of the phantom consists of a one voxel thick layer.

\subsubsection{Simulation setup}
The phantom is centered in the origin of the world coordinate system, and the particle source is placed in \SI{800}{\mm} distance anteroposterior to the phantom, such that the prostate lies approximately in the center of the photon beam.
The particle source radius is \SI{0.6}{\mm} and collimated to cover the same field as the manual dose measurements, resulting in aperture angles of \SI{7.02}{\degree} and \SI{9.04}{\degree}.
However, the resulting field size is indifferent, since the automatic exposure control is not active in the real-world experimental setup.
Emitted photon vertices are sampled using cosine-weighting to obtain homogeneous fluence with respect to a sphere surface.
The underlying energy spectrum of the photon shower is modeled considering a tungsten anode, \SI{70}{\kilo\volt} peak voltage, and \SI{2.7}{\mm} aluminum self-filtration using Boone's algorithm \cite{Boone:97}. 
% Particle interactions that may occur at the given energy spectrum are considered, including the photoelectric effect, Ray\-leigh scattering, and Compton scattering for photons and ionization and Bremsstrahlung for electrons. 
% All processes are modeled adhering to the Livermore model for low energy physics \cite{Perkins:91-1,Perkins:91-2,Cullen:97}. 
% Primary photons and secondary particles are tracked until their associated kinetic energy in consumed completely to satisfy energy preservation and assure accurate results.
\changed{M1.2}{\link{R1.2}}{Photon and electron interactions follow the standard low-energy electromagnetic physics defined in the Geant4 kernel (option 4), which is mainly based on the Livermore \cite{Perkins:91-1,Perkins:91-2,Cullen:97} and Penelope \cite{Sempau:97} models. The same Geant4 physics implementation was used in TG-195 \cite{Sechopoulos:15}.}

To obtain stable dose distributions, \num{10 x e8} primary photon histories are simulated. 
Dose distributions are scored with respect to the energy dose $D$ absorbed by each voxel measured in \si{\gray}. 
The simulation is carried out in batches of \num{e8} primaries in order to bring variance to the initial random seed and to split the computation to several nodes of the high performance computing (HPC) cluster. 
Each batch computation lasts on average \SI{3.5}{\hour}; however multiple batches are processed in parallel. 
The resulting dose distributions have the same resolution as the associated phantom volume. 
The simulated dose distributions are calibrated, such that the simulated primary and experimentally measured mean air kerma are equal.
This conversion is valid, since electric or collision kerma and absorbed dose are equivalent when charged particle equilibrium is present, which can be assumed for air inside of the primary photon beam.

Eventually, $H_T$ is calculated by averaging the dose absorbed in all voxels associated with a monitored organ.
In addition, using MC dose estimation, maximum dose values can be reported, which are of major interest regarding large organs that are irradiated unevenly.  
Due to limited MPs being available, this is not possible in experimental studies.
For large organs, such as skin and colon, only voxels that are irradiated directly are considered to obtain results consistent with the conducted dose measurements.

\section{Results}
\label{sec:results}

\subsection{Dose measurements}

% Dose Measurements
\begin{table*}
	\centering
	\caption{Measured organ-equivalent dose $H_T$ values from X-ray exposure of the central section interface and \SI{\pm 25}{\mm} offset as shown in Figure \ref{fig:setup-interface}, \ref{fig:setup-interface+}, and \ref{fig:setup-interface-}.  \changedtext{Data is shown as mean and standard deviation of all dosimeter readings per organ of all acquisitions per source/detector position.}}
    \label{tab:measurements-complete}
    \resizebox{\textwidth}{!}{
 	\begin{tabular}{lccccccccc}
    	\toprule
        \multirow{2}{*}{Organ} 
            & \multicolumn{2}{c}{$H_T$ [mGy] at \SI{0}{\mm}} & \multirow{2}{*}{$\dfrac{\text{AR}}{\text{CA}}$}
            & \multicolumn{2}{c}{$H_T$ [mGy] at \SI{+25}{\mm}} & \multirow{2}{*}{$\dfrac{\text{AR}}{\text{CA}}$} 
            & \multicolumn{2}{c}{$H_T$ [mGy] at \SI{-25}{\mm}} & \multirow{2}{*}{$\dfrac{\text{AR}}{\text{CA}}$} \\
        \cmidrule{2-3}\cmidrule{5-6}\cmidrule{8-9}
        & CA & AR & & CA & AR & & CA & AR  \\
        \midrule
        Prostate & 4.73 $\pm$ 0.36 & 4.64 $\pm$ 0.30 & 0.98 & 7.91 $\pm$ 0.48 & 8.03 $\pm$ 0.73 & 1.02 & 3.98 $\pm$ 0.07 & 3.76 $\pm$ 0.47 & 0.94 \\
        Bladder  & 2.31 $\pm$ 0.29 & 2.50 $\pm$ 0.69 & 1.08 & 3.07 $\pm$ 0.90 & 3.10 $\pm$ 1.51 & 1.01 & 2.44 $\pm$ 0.27 & 2.27 $\pm$ 0.46 & 0.93 \\
        Colon    & 5.51 $\pm$ 1.46 & 5.86 $\pm$ 1.24 & 1.06 & 11.37 $\pm$ 6.20 & 13.75 $\pm$ 7.78 & 1.21 & 6.71 $\pm$ 1.00 & 6.69 $\pm$ 0.75 & 1.00 \\
        Skin     & 36.66 $\pm$ 2.40 & 40.29 $\pm$ 4.49 & 1.10 & 34.97 $\pm$ 2.70 & 39.49 $\pm$ 4.35 & 1.13 & 36.69 $\pm$ 2.92 & 42.06 $\pm$ 4.62 & 1.15 \\
        \midrule
        Exit     & 0.54 $\pm$ 0.27 & 0.55 $\pm$ 0.09 & 1.02 & 0.77 $\pm$ 0.11 & 0.48 $\pm$ 0.06 & 0.62 & 0.51 $\pm$ 0.12 & 0.63 $\pm$ 0.14 & 1.24 \\
        \bottomrule
	\end{tabular}
	}
\end{table*}

Table \ref{tab:measurements-complete} summarizes the experimental organ-equivalent dose $H_T$ values for prostate, bladder, colon, and skin for the central section interface and \SI{\pm 25}{\mm} offset for both anthropomorphic phantoms as well as the corresponding exit dose. 
Overall, a high agreement between CA and AR phantoms can be reported for the majority of internal organs, except the colon, within \SIrange{92}{100}{\percent} accordance.
The measured $H_T$ values for the colon differ by \SI{21}{\percent} between both phantoms in one case.
Regarding skin dose, deviations of \SI{13}{\percent} in average are observable, where the AR phantom overestimates the skin entrance dose in comparison to the CA phantom.
The exit dose shows higher variation, however, the measured values are in the same order of magnitude.

\subsection{Impact of phantom placement}

% Impact of Phantom Placement
\begin{table}
	\centering
	\caption{Deviation ratios in organ-equivalent dose $H_T$ when moving the X-ray field \SI{\pm 25}{\mm} above or below with respect to the original view (see Table \ref{tab:measurements-complete}). The mean relative error is \SI{46}{\percent}.}
    \label{tab:placement}
	\begin{tabular}{@{\extracolsep{6pt}}lccccc@{}}
    	\toprule
        \multirow{2}{*}{Organ} & \multicolumn{2}{c}{$H_T$ ratio (CA)} & \multicolumn{2}{c}{$H_T$ ratio (AR)}\\
        \cmidrule{2-3}\cmidrule{4-5}
        & \SI{+25}{\mm} & \SI{-25}{\mm} & \SI{+25}{\mm} & \SI{-25}{\mm} \\
        \midrule
        Prostate & 1.67 & 0.84 & 1.73 & 0.81 \\
        Bladder & 1.33 & 1.06 & 1.24 & 0.91 \\
        Colon & 2.06 & 1.22 & 2.35 & 1.41 \\
        Skin & 0.95 & 1.00 & 1.04 & 0.98 \\
        \midrule
        Exit & 1.43 & 0.94 & 0.87 & 1.15 \\
        \bottomrule
    	
	\end{tabular}
\end{table}

To quantify the uncertainty induced by phantom (and patient) positioning, the influence of marginal displacements along the \longaxis{} axis by \SI{\pm 25}{\mm} has been investigated.
To this end, the measured $H_T$ values for \SI{\pm 25}{\mm} translation are set in relation to measured $H_T$ values for the original position and the resulting ratios are listed in Table \ref{tab:placement}. 
The impact of displacement of the field of view in both directions is of similar magnitude for both phantoms. 
However, translation by \SI{+25}{\mm} leads to deviations up to \SIrange{24}{73}{\percent} regarding the mostly irradiated organs prostate and bladder, while translation in the opposite direction produces modest deviation ranging from \SIrange{6}{19}{\percent}.
The skin dose is almost constant for both phantoms and \SI{\pm 25}{\mm} translation with a maximum deviation of \SI{5}{\percent}. 
Taking all monitored organs into account, the translation-induced deviation from the original field of view is \SI{46}{\percent} in average, however, with values ranging from \SIrange{6}{135}{\percent}.
The exit dose varies by up to \SI{43}{\percent}.

\subsection{Monte Carlo simulation}

Figure \ref{fig:mc-reference-dose} to \ref{fig:mc-water-dose} show an exemplary axial slice of the MC simulated dose distributions inside the digital Golem phantom at the height of the prostate for all four material mappings.
The calculated organ-equivalent dose values are juxtaposed to the associated measurements in Table \ref{tab:measurements-simulation}.
In general, the $H_T$ values estimated from MC simulations are in good agreement with the measured ones for both physical phantoms concerning the prostate, where an accuracy of \SIrange{95}{99}{\percent} is achieved.
However, regarding the colon of both and the bladder of the AR phantom, respectively, deviations of \SIrange{12}{22}{\percent} are observable.
The exit dose is underestimated by \SIrange{13}{22}{\percent}.
Taking the entrance skin dose into account, divergence of up to \SI{35}{\percent} between measured and simulated $H_T$ values for the AR phantom are conspicuous.
\changed{M2.6}{\link{R2.6}}{Therefore we reiterated the corresponding MC simulations and replaced the formerly used soft-tissue-equivalent material with water, which is commonly used to approximate soft tissue.
These novel MC simulations} (ARW; Figure \ref{fig:mc-water-dose}) resulted in a reduction of the deviation regarding skin dose to \SI{11}{\percent}, while internal $H_T$ values still correspond well to the measurements. 
% MC simulation considering water as soft-tissue-equivalent material (ARW; Figure \ref{fig:mc-water-dose}) resulted in a reduction of the deviation regarding skin dose to \SI{11}{\percent}, while internal $H_T$ values still correspond well to the measurements. 

\subsection{Impact of material composition}

% Monte Carlo Simulation
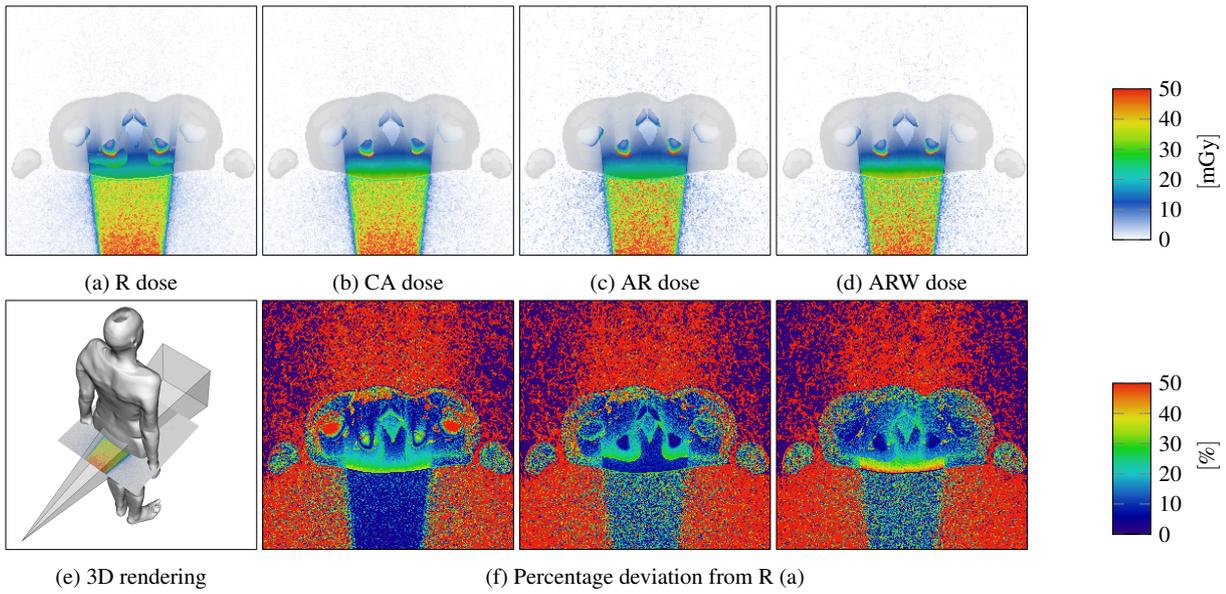
\begin{figure*}
	\centering
    \subcaptionbox{R dose\label{fig:mc-reference-dose}}
    {\includegraphics[frame,width=0.19\linewidth]{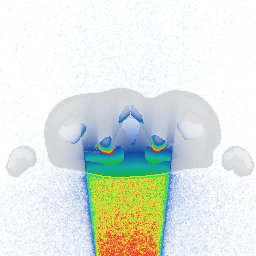}}
    \subcaptionbox{CA dose\label{fig:mc-atom-dose}}
    {\includegraphics[frame,width=0.19\linewidth]{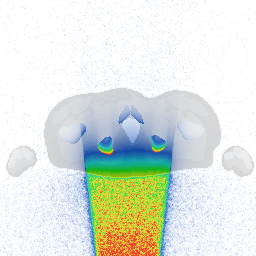}}
    \subcaptionbox{AR dose\label{fig:mc-rando-dose}}
    {\includegraphics[frame,width=0.19\linewidth]{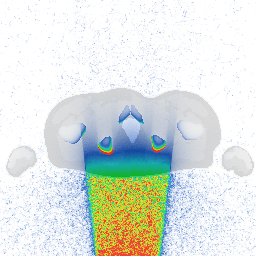}}
    \subcaptionbox{ARW dose\label{fig:mc-water-dose}}
    {\includegraphics[frame,width=0.19\linewidth]{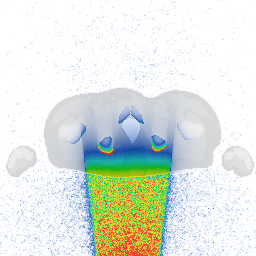}}
    \begin{subfigure}[b]{0.19\linewidth}
    	\centering
    	\begin{tikzpicture}
				\begin{axis}[
    				hide axis,
    				scale only axis,
    				height=0pt,
    				width=0pt,
    				colormap name=dosemap,
    				colorbar,
    				point meta min = 0,
    				point meta max = 50,
    				colorbar style={
    				    height=2cm,  % 3cm,
                        ytick={0, 10, 20, 30, 40, 50},
                        y unit=\si{\milli\gray}
                    },
                    colorbar/width=0.5cm  % 0.75cm
    			]
    				\addplot [draw=none] coordinates {(0,0)};
				\end{axis}
		\end{tikzpicture}
    \end{subfigure}
    
    \subcaptionbox{3D rendering\label{fig:mc-3d}}
    {\includegraphics[frame,width=0.19\linewidth]{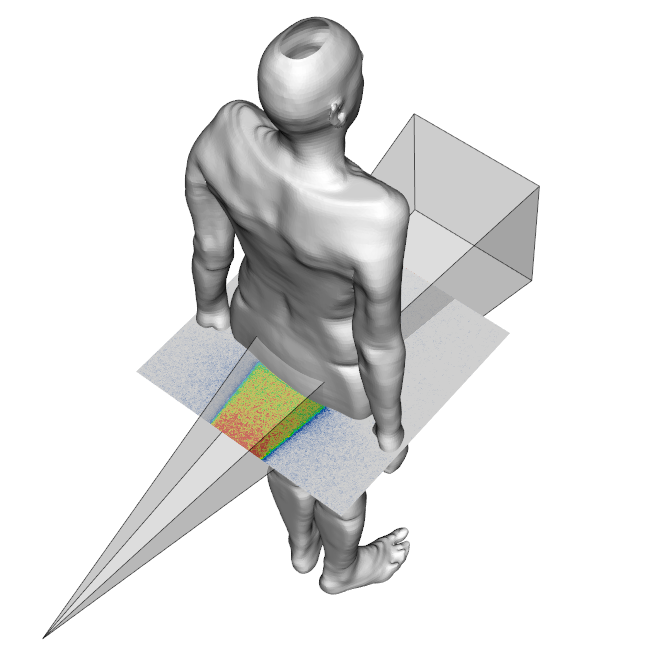}}
    \subcaptionbox{Percentage deviation from R (a)\label{fig:mc-error}}
    {\includegraphics[frame,width=0.19\linewidth]{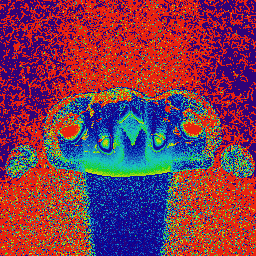}
    \includegraphics[frame,width=0.19\linewidth]{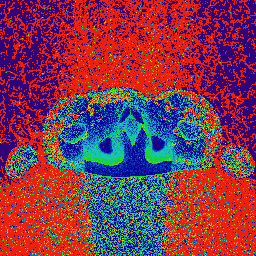}
    \includegraphics[frame,width=0.19\linewidth]{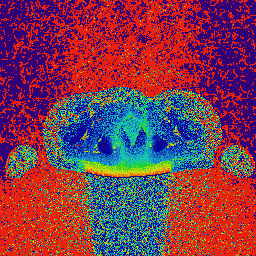}}
    \begin{subfigure}[b]{0.19\linewidth}
    	\centering
    	\begin{tikzpicture}
				\begin{axis}[
    				hide axis,
    				scale only axis,
    				height=0pt,
    				width=0pt,
    				colormap name=physicsmap,
    				colorbar,
    				point meta min = 0,
    				point meta max = 50,
    				colorbar style={
    				    height=2cm,  % 3cm, 
    				    ytick={0, 10, 20, 30, 40, 50},
                        y unit=\si{\percent}
                    },
                    colorbar/width=0.5cm  % 0.75cm
    			]
    				\addplot [draw=none] coordinates {(0,0)};
				\end{axis}
			\end{tikzpicture}
    \end{subfigure}
    \caption{(a)-(d) Axial slice of the digital Golem phantom overlaid with the deposited X-ray dose distribution associated with each set of material properties. The prostate is located in the center of both the phantom and the X-ray beam. (f) Corresponding percentage deviation maps with respect to the reference R. (e) Spatial relation between dose maps and phantom. 
    }
    \label{fig:mc}
\end{figure*}

% Monte Carlo Simulation
\begin{table*}
    \centering
    \caption{
    Comparison of measured and MC simulated organ-equivalent dose $H_T$ for all considered phantoms from irradiation of the central section interface as shown in Figure \ref{fig:setup-interface}. 
    $H_T$ is given in terms of mean measurement and standard deviation.
    The standard deviation associated with the MC simulations derives from multiple voxels contributing to the organ dose whereas the standard deviation associated with the measurements derives from multiple MPs and repetitions. 
    \changedtext{The uncertainty of the MOSFET probes is not included.}
    For the ARW phantom, the measured values of the AR phantom are assumed.
    }
    \label{tab:measurements-simulation}
    \resizebox{\textwidth}{!}{
    \begin{tabular}{lcccccccccc}
        \toprule
        \multirow{2}{*}{Organ} 
            & \multicolumn{2}{c}{$H_T$ [\si{\milli\gray}] CA} & \multirow{2}{*}{$\dfrac{\text{Sim.}}{\text{Meas.}}$} 
            & \multicolumn{2}{c}{$H_T$ [\si{\milli\gray}] AR} & \multirow{2}{*}{$\dfrac{\text{Sim.}}{\text{Meas.}}$} 
            & \multicolumn{2}{c}{$H_T$ [\si{\milli\gray}] ARW} & \multirow{2}{*}{$\dfrac{\text{Sim.}}{\text{Meas.}}$}\\
        \cmidrule{2-3}\cmidrule{5-6}\cmidrule{8-9}
        & Measured & Simulated & & Measured & Simulated & & Measured & Simulated  \\
        \midrule
        Prostate & 4.73 $\pm$ 0.36 & 4.49 $\pm$ 1.44 & 0.95 & 4.64 $\pm$ 0.30 & 4.60 $\pm$ 1.49 & 0.99   & 4.64 $\pm$ 0.30 & 4.50 $\pm$ 1.54 & 0.97 \\
        Bladder & 2.31 $\pm$ 0.29 & 2.17 $\pm$ 0.96 & 0.94  & 2.50 $\pm$ 0.69 & 2.18 $\pm$ 1.00 & 0.87   & 2.50 $\pm$ 0.69  & 2.11 $\pm$ 0.97 & 0.84 \\
        Colon & 5.51 $\pm$ 1.46 & 6.70 $\pm$ 1.07 & 1.22    & 5.86 $\pm$ 1.24 & 6.58 $\pm$ 0.98 & 1.12   & 5.86 $\pm$ 1.24    & 6.58 $\pm$ 1.15 & 1.12 \\
        Skin & 36.66 $\pm$ 2.4 & 33.66 $\pm$ 1.69 & 0.92    & 40.29 $\pm$ 4.49 & 26.21 $\pm$ 1.43 & 0.65 & 40.29 $\pm$ 4.49   & 36.70 $\pm$ 1.89 & 0.89 \\
        \midrule
        Exit & 0.54 $\pm$ 0.27 & 0.47 $\pm$ 0.18 & 0.87 & 0.55 $\pm$ 0.09 & 0.47 $\pm$ 0.21 & 0.85 & 0.55 $\pm$ 0.09 & 0.43 $\pm$ 0.20 & 0.78 \\
        \bottomrule
    \end{tabular}
    }
\end{table*}

% Impact of Material Composition
\begin{table}
	\centering
	\caption{
	Comparison of simulated organ-equivalent dose $H_T$ and effective dose $E$ for the CA, AR, ARW (see Table \ref{tab:measurements-simulation}), and R phantom. 
	$H_T$ is given in terms of mean organ dose over all directly irradiated voxels.
	}
	\label{tab:simulation}
 	\begin{tabular}{@{\extracolsep{6pt}}lcccc@{}}
    	\toprule
        \multirow{2}{*}{Organ} & $H_T$ [mGy] & \multicolumn{3}{c}{Ratio}\\
        \cmidrule{2-2}\cmidrule{3-5}
        & R & CA / R & AR / R & ARW / R \\
        \midrule
        Prostate & 5.61 & 0.80 & 0.82 & 0.80 \\
        Bladder & 2.47 & 0.88 & 0.88 & 0.85 \\
        Colon & 8.02 & 0.84 & 0.82 & 0.82 \\
        Skin & 36.50 & 0.92 & 0.72 & 1.01 \\
        \midrule
        Exit & 0.50 & 0.94 & 0.94 & 0.86 \\
        \midrule
        \midrule
        Effective & \SI{0.57}{\milli\sievert} & 0.89 & 0.95 & 0.88 \\
        \bottomrule
	\end{tabular}
\end{table}

The impact of tissue-equivalent material composition on the dose deposition is investigated based on the results of the MC simulations.
Table \ref{tab:simulation} lists the organ-equivalent dose values $H_T$ of the computational R, CA, AR, and ARW phantoms as well as the ratio between $H_T$ values of the artificial anthropomorphic phantoms and the reference. 
The usage of artificial material compounds leads to an underestimation of absorbed dose by \SI{8}{\percent} for the CA phantom and \SI{28}{\percent} for the AR phantom regarding average skin dose.
The highest accordance concerning skin dose was found with the ARW phantom, where only a deviation of \SI{1}{\percent} occurred. 
However, the organ dose deposited in all internal organs is underestimated with a deviation of \SIrange{12}{20}{\percent}, with all phantom materials performing comparably.
Combining all organs, the average deviation for the CA phantom is \SI{14}{\percent}, \SI{19}{\percent} for the AR phantom, and \SI{14}{\percent} for the ARW phantom, respectively. 
Concerning effective dose, deviations of \SIrange{5}{12}{\percent} with respect to the reference R can be reported.

The associated deviation maps for an exemplary axial slice of the three artificially composed phantoms, shown in Figure \ref{fig:mc-error}, support these findings. 
Note that noisy distributed, large deviations in the air and outside of the primary X-ray beam are mainly due to the overall statistical uncertainty involved when scoring dose distributions directly.
Since dose is deposited by secondary electrons, it can only be scored when an actual interaction is sampled, thus the evaluation of the here carried out MC simulations only refers to directly irradiated body parts.
Although there are individual regions or tissue types where the dose distribution is estimated accurately with respect to the reference R, there are also large areas where the deviation is significantly greater than \SI{15}{\percent}.

\section{Discussion}
\label{sec:discussion}
To prevent deterministic and stochastic radiation-induced damage to the  patient, interventional X-ray exposure should be as low as reasonably achievable.
To obtain measures for the expected patient dose in certain procedures, phantom studies using dedicated anthropomorphic dosimetry phantoms are in general conducted.
This approach, however, introduces multiple caveats concerning FGIs, which are in general less predictable than purely diagnostic procedures due to possible complications and patient specificities.
Besides uncertainties associated with the FGI workflow and patient geometry and placement, the artificial material compounds used in modern anthropomorphic phantoms often represent an approximation of mixtures of human tissues, e.g. an averaged tissue-equivalent material for soft, adipose, and muscular tissue.
While resulting in adequate average linear attenuation coefficients, these mixture materials combined with a not patient-specific geometry, might not correctly account for local dose deposition.
However, especially with FGIs concerning the improvement of life quality of otherwise healthy patients, such as in PAE, accurate estimation of patient dose is necessary to correctly assess the cost-benefit relation.

Here, we investigated certain sources of uncertainty correlated to the determination of organ-equivalent dose values $H_T$ in FGIs on the example of one possible 2D imaging situation during a PAE procedure.
Therefore, we used two renowned anthropomorphic dosimetry phantoms in combination with high-sensitivity MOSFET dosimeters to measure and subsequently calculate the average dose deposited in internal organs as well as the skin entrance dose.
To guarantee sensitive MOSFET measurements, we applied unusual high radiation exposure per measurement, leading to relatively high absolute dose values.
In addition, we employed an MC radiation transport code to reconstruct the experiment digitally to obtain dense dose distributions. This allows, first, to validate the simulation against the experiment, and second to determine the influence of material composition and limited sampling points of the phantoms on the estimated experimental organ doses.

\changed{M1.3}{\link{R1.3}}{Voxel size may influence simulation output. However, since we mainly compare the simulation results among each other, derived conclusions concerning the impact of material composition are not affected.
In addition, the large slice thickness of \SI{5}{\cm} of the physical phantoms and the arrangement of holes in a rectangular grid, only allows for a sparse sampling of measurement points.
Therefore it might be questionable, whether a finer voxel size would yield significant improvements.
Regarding, the comparison between simulated and measured dose values, however, the influence of voxel size could be evaluated in more depth in future studies.}

Although both investigated anthropomorphic phantoms vary considerably in the composition of tissue-equivalent compounds, the dose values derived from MOSFET dosimeter measurements were in accordance with\-in a maximum deviation of \SI{8}{\percent} in the majority of organs and imaging positions. 
However, skin and colon dose showed higher variations between the two phantoms.

\changed{M1.5}{\link{R1.5}}{The discrepancy concerning bladder and exit dose values likely stems from the overall low dose values in these regions leading to high uncertainties of the MOSFET probes.}
Dose deposited in the colon, showed a variation up to \SI{22}{\percent}, which could not be accounted for by the slightly different equipping of the phantoms. Therefore, we conclude, that the limited amount of dosimetry measurement points are for an organ with a complex geometry, such as the colon, not sufficient to correctly estimate organ dose.
This observation is also substantiated taking the \changedtext{remaining} MC simulated dose \changedtext{values} into account, as they are in overall agreement with the measurements, only the colon shows a deviation up to \SI{22}{\percent} from the experiments. 
To this end, the importance of developing and incorporating patient-specific digital twins is substantiated in order to provide most accurate dose assessment tools.

Regarding skin dose values, the MC simulations yield ambiguous results.
On the one hand, the average skin dose of the digital CA phantom can be calculated to be in \SI{92}{\percent} accordance with the physical measurements.
On the other hand, a divergence of \SI{35}{\percent} is reported for the digital AR phantom, which leads to the assumption that there is a discrepancy between the digital material model of its soft-tissue-equivalent and the actual physical plastic.
Reiterating the concerning MC simulation with water instead of the soft-tissue-equivalent yields comparable results for the internal, while the deviation in skin dose is considerably lower with \SI{11}{\percent}.
\changed{M2.8}{\link{R2.8}}{Whether these findings concerning the several decades old AR phantom might be due to age- and usage-related material wear, could be investigated in the future.}
% These findings concerning the AR phantom might be due to age- and usage-related material wear, since the used phantom is in use for several decades.

Additionally, the material composition of tissue-mimicking phantom materials influences the determined organ doses. 
The computationally determined organ doses differ additionally from the reference by up to \SI{28}{\percent}. 
This is for a clinical decision specifically relevant, as all the dose values of the artificial materials underestimate organ doses compared to the human reference material. 
This is further substantiated regarding effective dose, where underestimation by \SIrange{5}{12}{\percent} could be observed.
\changed{M2.7.1}{\link{R2.7}}{To allow for reliable assessment of measured dose values in clinically motivated phantom studies, this underestimation has to be taken into account properly.}

Finally, we explored the influence of geometric uncertainties, such as patient or X-ray source positioning, on the organ dose values.
Therefore we measured dose values in the two anthropomorphic phantoms with respect to three distinct positions along the \longaxis{} axis of the phantoms, uniformly spread over \SI{50}{\mm}.
We considered the central measurement as baseline and regarded the ratio of the measurements translated by $\SI{\pm 25}{\mm}$ to this reference.
The displacement-induced deviations range from \SIrange{6}{135}{\percent} for the internal organs, only skin dose remains relatively constant. 
\changed{M2.4}{\link{R2.4}}{Regarding the internal organs, deviations are caused by the altering organ coverage by the primary X-ray field, thereby changing the scatter-to-primary ratio and the shielding by bones.}
This leads to the conclusion, that patient positioning plays a crucial role to correctly assess and manage radiation exposure, as small deviations may have a great impact on organ dose. 
\changed{M3.3.2}{\link{R3.3}}{Future studies might also investigate the impact of angular variations, since, regarding the case of PAE for instance, ipsilateral angulation of \SIrange{25}{35}{\degree} is a common choice.}
\changed{M2.3}{\link{R2.3}}{Also comparing the reference voxel phantom to CT scans of both physical phantoms might allow for a further in depth analysis.}

It has to be additionally considered, that in a clinical procedure the geometric uncertainty comprises besides positioning  also the geometry of the patient, which may vary in a much greater extent. 
Also, slightly divergent projection geometries applied in the procedure invalidate prospectively determined reference dose values. 
Therefore, it appears inevitable to introduce computer-assisted methods to (a) model or customize a digital twin of the patient and (b) support the medical staff in registering this patient model to the actual patient, in order to enable online dose simulation or calculation.
Online dose simulation could also yield 3D dose distributions and therefore indicate internal areas of maximum radiation exposure instead of single reference $E$ and $H_T$ values or sparsely sampled MPs of direct dosimeters.

Since it is generally challenging to integrate MC calculations with experimental dosimetry, the reported deviation of \SIrange{1}{5}{\percent} concerning the prostate, and \SIrange{6}{13}{\percent} concerning the bladder, respectively, can be considered negligible and might be due to stochastic fluctuations.
Colon and skin deviations seem to be linked to geometric limitations of the experimental setup, as discussed before. 
It is further emphasized, that the simulated phantom geometry does not match the physical anthropomorphic phantoms exactly, but is modeled to resemble an adult reference male. 
Additionally, it has to be considered, that the accuracy of the dosimeters is limited due to their exposure low dose limit of \SI{1.69}{mGy} and statistically insignificant energy dependency \cite{koivisto2015characterization}. 
Therefore, and due to comparable validation accuracy to other computational approaches \cite{marshall2018organ,khodadadegan2013validation,golikov2017comparative}, we consider our simulation validated.

\section{Conclusion}
\label{sec:conclusion}
We demonstrated the impact of using different phantoms, with focus on tissue material model, and phantom positioning on the outcome of prospective phantom dosimetry studies. 
Considerable deviations in organ dose estimation due to uncertainties associated with experimental dosimetry were found, implying the need for individual patient-specific dose distribution estimation, e.g. using MC simulation. 
\changed{M2.7.2}{\link{R2.7}}{In addition, the found deviations deriving from artificial tissue-equivalent materials underline the need for materials optimized not only in terms of linear attenuation but also energy absorption.}
Future studies will combine \changedtext{the used} simulation setup with advanced patient modeling methods as well as robust algorithms to register and continuously adjust this digital twin to the actual patient.

\section*{Compliance with Ethical Standards}
\textbf{Conflict of Interest~~} A. Maier has no conflict of interest to declare. P. Roser is funded by the Erlangen Graduate School in Advanced Optical Technologies. X. Zhong is now with the Siemens Healthcare GmbH. During the work on this research, X. Zhong has not been affiliated with Siemens. P. Ochs is now with the Nuremberg Institute of Technology Georg Simon Ohm. During the work on this research, P. Ochs has been employee of the Siemens Healthcare GmbH. A. Birkhold, E. Stepina, M. Kowarschik, and R. Fahrig are employees of the Siemens Healthcare GmbH.

~\\
\textbf{Human and Animal participants~~} This article does not contain any studies with human participants or animals performed by any of the authors.

~\\
\textbf{Disclaimer~~} The concepts and information presented in this article are based on research and are not commercially available.

% \bibliographystyle{unsrt}
% \bibliography{references.bib}   % name your BibTeX data base

\begin{thebibliography}{10}

\bibitem{Falco:18}
MD~Falco, S~Masala, M~Stefanini, P~Bagal\`a, D~Morosetti, E~Calabria,
  A~Tonnetti, and G~Verona-Rinati.
\newblock Effective-dose estimation in interventional radiological procedures.
\newblock {\em Radiol Phys Technol}, 11:149--155, 2018.

\bibitem{Balter:10}
S~Balter, JW~Hopewell, DL~Miller, LK~Wagner, and MJ~Zelefsky.
\newblock Fluoroscopically guided interventional procedures: a review of
  radiation effects on patients' skin and hair.
\newblock {\em Radiology}, 254(2):326--341, 2010.

\bibitem{Wagner:99}
LK~Wagner, MD~McNeese, MV~Marx, and EL~Siegel.
\newblock Severe skin reactions from interventional fluoroscopy: Case report
  and review of the literature.
\newblock {\em Radiology}, 213(3):773--776, 1999.

\bibitem{Koenig:01}
TR~Koenig, D~Wolff, FA~Mettler, and LK~Wagner.
\newblock Skin injuries from fluoroscopically guided procedures: Part 1,
  characteristics of radiation injury.
\newblock {\em AJR Am J Roentgenol}, 177(1):3--11, 2001.

\bibitem{Dincan:02}
M~D'Incan, H~Roger, J~Gabrillargues, S~Mansard, S~Parent, J~Chazal, B~Irthum,
  and P~Souteyrand.
\newblock Radiation-induced temporary hair loss after endovascular embolization
  of the cerebral arteries: Six cases.
\newblock {\em Ann {D}ermatol {V}enereol.}, 129(5):703--706, 2002.

\bibitem{Balter:06}
S~Balter.
\newblock Methods for measuring fluoroscopic skin dose.
\newblock {\em Pediatr Radiol}, 36(2):136--140, 2006.

\bibitem{Mattsson:16}
S~Mattsson.
\newblock Need for individual cancer risk estimates in x-ray and nuclear
  medicine imaging.
\newblock {\em Radiat Prot Dosimetry}, 169(1--4):11--16, 2016.

\bibitem{Carnevale:10}
FC~Carnevale, AA~Antunes, JM~da~Motta Leal~Filho, LM~de~Oliveira~Cerri,
  RH~Baroni, ASZ Marcelino, GC~Freire, AM~Moreira, M~Srougi, and GG~Cerri.
\newblock Prostatic artery embolization as a primary treatment for benign
  prostatic hyperplasia: Preliminary results in two patients.
\newblock {\em Cardiovasc Intervent Radiol}, 33:355--361, 2010.

\bibitem{Bagla:13}
S~Bagla, CP~Martin, A~van Breda, MJ~Sheridan, KM~Sterling, D~Papadouris,
  KS~Rholl, JB~Smirniotopoulos, and A~van Breda.
\newblock Early results from a united states trial of prostatic artery
  embolization in the treatment of benign prostatic hyperplasia.
\newblock {\em J Vasc Interv Radiol}, 25(1):47--52, 2013.

\bibitem{Christidis:18}
D~Christidis, E~Clarebrough, V~Ly, M~Perera, H~Woo, N~Lawrentschuk, and
  D~Bolton.
\newblock Prostatic artery embolization for benign prostatic obstruction:
  Assessment of safety and efficacy.
\newblock {\em World J Urol}, 36:575--584, 2018.

\bibitem{Garzon:16}
WJ~Garz\'on, G~Andrade, F~Dubourcq, DG~Abud, M~Bredow, HJ~Khoury, and R~Kramer.
\newblock Prostatic artery embolization: radiation exposure to patients and
  staff.
\newblock {\em J Radiol Prot}, 36:246--254, 2016.

\bibitem{Laborda:15}
A~Laborda, AM~De~Assis, I~Ioakeim, M~S{\'a}nchez-Ballest{\'i}n, FC~Carnevale,
  and MA~De~Gregorio.
\newblock Radiodermitis after prostatic artery embolization: Case report and
  review of the literature.
\newblock {\em Cardiovasc Intervent Radiol}, 38:755--759, 2015.

\bibitem{Andrade:17}
G~Andrade, HJ~Khoury, WJ~Garz\'on, F~Dubourcq, M~Bredow, LM~Monsignore, and
  DG~Abud.
\newblock Radiation exposure of patients and interventional radiologists during
  prostatic artery embolization: a prospective single-operator study.
\newblock {\em J Vasc Interv Radiol}, 28(4):517--521, 2017.

\bibitem{Tanaka:17}
M~Tanaka, E~Lacayo, J~Katrivesis, J~Spies, and A~Kim.
\newblock Radiation doses in prostatic artery embolization for benign prostatic
  hypertrophy: A single-institution series and meta-analysis.
\newblock {\em J Vasc Interv Radiol}, 28(2, Supplement):S149, 2017.

\bibitem{Gao:14}
Y~Gao, Y~Huang, R~Zhang, Y~Yang, Q~Zhang, M~Hou, and Y~Wang.
\newblock Benign prostatic hyperplasia: Prostatic arterial embolization versus
  transurethral resection of the prostate--{A} prospective, randomized, and
  controlled clinical trial.
\newblock {\em Radiology}, 270(3):920--928, 2014.

\bibitem{Pisco:16}
JM~Pisco, T~Bilhim, LC~Pinheiro, L~Fernandes, J~Pereira, NV~Costa, M~Duarte,
  and AG~Oliveira.
\newblock Medium- and long-term outcome of prostate artery embolization for
  patients with benign prostatic hyperplasia: Results in 630 patients.
\newblock {\em J Vasc Interv Radiol}, 27(8):1115--1122, 2016.

\bibitem{Badal:09}
A~Badal and A~Badano.
\newblock Accelerating {Monte Carlo} simulations of photon transport in a
  voxelized geometry using a massively parallel graphics processing unit.
\newblock {\em Med Phys}, 36(11):4878--4880, 2009.

\bibitem{Bert:13}
J~Bert, H~Perez-Ponce, Z~El Bitar, S~Jan, Y~Boursier, D~Vintache, A~Bonissent,
  C~Morel, D~Brasse, and D~Visvikis.
\newblock Geant4-based {Monte Carlo} simulations on {GPU} for medical
  applications.
\newblock {\em Phys Med Biol}, 58(16):5593--5611, 2013.

\bibitem{Shrimpton:81}
PC~Shrimpton, BF~Wall, and ES~Fisher.
\newblock The tissue-equivalence of the {A}lderson {R}ando anthropomorphic
  phantom for x-rays of diagnostic qualities.
\newblock {\em Phys Med Biol}, 26(1):133--139, 1981.

\bibitem{White:78}
DR~White.
\newblock Tissue substitutes in experimental radiation physics.
\newblock {\em Med Phys}, 5(6):467--479, 1978.

\bibitem{Sechopoulos:15}
I~Sechopoulos and ESM~Ali and A~Badal and A~Badano and JM~Boone and IS~Kyprianou and E~Mainegra-Hing and KL~McMillan and MF~McNitt-Gray and DWO~Rogers and E~Samei and AC~Turner.
\newblock Monte Carlo reference data sets for imaging research: Executive summary of the report of AAPM Research Committee Task Group 195.
\newblock {\em Med Phys}, 42(10):5679--5691, 2015.

\bibitem{Sempau:97}
J~Sempau, E~Acosta, J~Baro, JM~Fern\'{a}ndez-Varea, and F~Salvat.
\newblock An algorithm for Monte Carlo simulation of coupled electron-photon transport.
\newblock {\em Nucl Instrum Meth B}, 132(3):377--390, 1997.

\bibitem{Agostinelli:03}
S~Agostinelli and J~Allison and K~Amako and J~Apostolakis and H~Araujo and P~Arce and M~Asai and D~Axen and S~Banerjee and G~Barrand and F~Behner and L~Bellagamba and J~Boudreau and L~Broglia and A~Brunengo and H~Burkhardt and S~Chauvie and J~Chuma and R~Chytracek and G~Cooperman and G~Cosmo and P~Degtyarenko and A~Dell'Acqua and G~Depaola and D~Dietrich and R~Enami and A~Feliciello and C~Ferguson and H~Fesefeldt and G~Folger and F~Foppiano and A~Forti and S~Garelli and S~Giani and R~Giannitrapani and D~Gibin and J.J~G\'{o}mez Cadenas and I~Gonz\'{a}lez and G~Gracia Abril and G~Greeniaus and W~Greiner and V~Grichine and A~Grossheim and S~Guatelli and P~Gumplinger and R~Hamatsu and K~Hashimoto and H~Hasui and A~Heikkinen and A~Howard and V~Ivanchenko and A~Johnson and FW~Jones and J~Kallenbach and N~Kanaya and M~Kawabata and Y~Kawabata and M~Kawaguti and S~Kelner and P~Kent and A~Kimura and T~Kodama and R~Kokoulin and M~Kossov and H~Kurashige and E~Lamanna and T~Lamp\'{e}n and V~Lara and V~Lefebure and F~Lei and M~Liendl and W~Lockman and F~Longo and S~Magni and M~Maire and E~Medernach and K~Minamimoto and P~Mora de Freitas and Y~Morita and K~Murakami and M~Nagamatu and R~Nartallo and P~Nieminen and T~Nishimura and K~Ohtsubo and M~Okamura and S~O'Neale and Y~Oohata and K~Paech and J~Perl and A~Pfeiffer and M.G~Pia and F~Ranjard and A~Rybin and S~Sadilov and E~Di Salvo and G~Santin and T~Sasaki and N~Savvas and Y~Sawada and S~Scherer and S~Sei and V~Sirotenko and D~Smith and N~Starkov and H~Stoecker and J~Sulkimo and M~Takahata and S~Tanaka and E~Tcherniaev and E~Safai Tehrani and M~Tropeano and P~Truscott and H~Uno and L~Urban and P~Urban and M~Verderi and A~Walkden and W~Wander and H~Weber and JP~Wellisch and T~Wenaus and DC~Williams and D~Wright and T~Yamada and H~Yoshida and D~Zschiesche.
\newblock Geant4--a simulation toolkit.
\newblock {\em Nucl Instrum Meth A}, 506(3):250--303, 2003.

\bibitem{Allison:16}
J~Allison and K~Amako and J~Apostolakis and P~Arce and M~Asai and T~Aso and E~Bagli and A~Bagulya and S~Banerjee and G~Barrand and BR~Beck and AG~Bogdanov and D~Brandt and JMC~Brown and H~Burkhardt and Ph~Canal and D~Cano-Ott and S~Chauvie and K~Cho and G.A.P~Cirrone and G~Cooperman and MA~Cort\'{e}s-Giraldo and G~Cosmo and G~Cuttone and G~Depaola and L~Desorgher and X~Dong and A~Dotti and VD~Elvira and G~Folger and Z~Francis and A~Galoyan and L~Garnier and M~Gayer and KL~Genser and VM~Grichine and S~Guatelli and P~Gu\`{e}ye and P~Gumplinger and AS~Howard and I~H\v{r}ivn\'{a}\v{c}ov\'{a} and S~Hwang and S~Incerti and A~Ivanchenko and VN~Ivanchenko and FW~Jones and SY~Jun and P~Kaitaniemi and N~Karakatsanis and M~Karamitros and M~Kelsey and A~Kimura and T~Koi and H~Kurashige and A~Lechner and SB~Lee and F~Longo and M~Maire and D~Mancusi and A~Mantero and E~Mendoza and B~Morgan and K~Murakami and T~Nikitina and L~Pandola and P~Paprocki and J~Perl and I~Petrovi\'{c} and MG~Pia and W~Pokorski and JM~Quesada and M~Raine and MA~Reis and A~Ribon and A~Risti\'{c} Fira and F~Romano and G~Russo and G~Santin and T~Sasaki and D~Sawkey and JI~Shin and II~Strakovsky and A~Taborda and S~Tanaka and B~Tom\'{e} and T~Toshito and H.N~Tran and P.R~Truscott and L~Urban and V~Uzhinsky and JM~Verbeke and M~Verderi and B.L~Wendt and H~Wenzel and DH~Wright and DM~Wright and T~Yamashita and J~Yarba and H~Yoshida.
\newblock Recent developments in Geant4.
\newblock {\em Nucl Instrum Meth A}, 835(Supplement C):186--225, 2016.

\bibitem{ICRP:110}
{International Commission on Radiological Protection}.
\newblock {ICRP} publication 110: Adult reference computation phantoms.
\newblock {\em Ann ICRP}, 39(2), 2009.

\bibitem{Boone:97}
JM~Boone and JA~Seibert.
\newblock An accurate method for computer-generating tungsten anode x-ray
  spectra from 30 to 140 kv.
\newblock {\em Med Phys}, 24(11):1661--1670, 1997.

\bibitem{Perkins:91-1}
ST~Perkins, DE~Cullen, and SM~Seltzer.
\newblock Tables and graphs of electron-interaction cross-sections from 10 {eV}
  to 100 {GeV} derived from the {LLNL Evaluated Electron Data Library}
  ({EEDL}), {Z}= 1-100.
\newblock Technical report, Lawrence Livermore National Laboratory, 1991.

\bibitem{Perkins:91-2}
ST~Perkins, DE~Cullen, MH~Chen, J~Rathkopf, J~Scofield, and JH~Hubbell.
\newblock Tables and graphs of atomic subshell and relaxation data derived from
  the {LLNL Evaluated Atomic Data Library} ({EADL}), {Z}= 1--100.
\newblock Technical report, Lawrence Livermore National Laboratory, 1991.

\bibitem{Cullen:97}
DE~Cullen, JH~Hubbell, and L~Kissel.
\newblock {EPDL}97: The evaluated photon data library, 97 version.
\newblock Technical report, Lawrence Livermore National Laboratory, 1997.

\bibitem{koivisto2015characterization}
JH~Koivisto, JE~Wolff, T~Kiljunen, D~Schulze, and M~Kortesniemi.
\newblock Characterization of {MOSFET} dosimeters for low-dose measurements in
  maxillofacial anthropomorphic phantoms.
\newblock {\em J Appl Clin Med Phys}, 16(4):266--278, 2015.

\bibitem{marshall2018organ}
EL~Marshall, D~Borrego, JC~Fudge, D~Rajderkar, and WE~Bolch.
\newblock Organ doses in pediatric patients undergoing cardiac-centered
  fluoroscopically-guided interventions: Comparison of three methods for
  computational phantom alignment.
\newblock {\em Med Phys}, 45(8):3926--3938, 2018.

\bibitem{khodadadegan2013validation}
Y~Khodadadegan, M~Zhang, W~Pavlicek, RG~Paden, B~Chong, EA~Huettl, BA~Schueler,
  KA~Fetterly, SG~Langer, and T~Wu.
\newblock Validation and initial clinical use of automatic peak skin dose
  localization with fluoroscopic and interventional procedures.
\newblock {\em Radiology}, 266(1):246--255, 2013.

\bibitem{golikov2017comparative}
V~Golikov, A~Barkovsky, E~Wallstr{\=o}m, and {\AA}~Cederblad.
\newblock A comparative study of organ doses assessment for patients undergoing
  conventional x-ray examinations: phantom experiments vs. calculations.
\newblock {\em Radiat Prot Dosimetry}, 178(2):223--234, 2017.

\end{thebibliography}

% \begin{thebibliography}{}


% Author, Article title, Journal, Volume, page numbers (year)
% Author, Book title, page numbers. Publisher, place (year)
% etc
% \end{thebibliography}

\end{document}